# All-optical nanoscale thermometry based on silicon-vacancy centers in detonation nanodiamonds

Masanori Fujiwara[1], Gaku Uchida[1], Izuru Ohki[1], Ming Liu[2], Akihiko Tsurui[2], Taro Yoshikawa[2], Masahiro Nishikawa[2], and Norikazu Mizuochi[1,3*]

[1] Institute for Chemical Research, Kyoto University, Gokasho, Uji, Kyoto 611-0011, Japan

[2] Innovation and Business Development Headquarters, Daicel Corporation, 1239, Shinzaike, Aboshi-ku, Himeji, Hyogo 671-1283, Japan

[3] Center for Spintronics Research Network, Kyoto University, Gokasho, Uji, Kyoto 611-0011, Japan

*Corresponding Author: Email: mizuochi@scl.kyoto-u.ac.jp

**Abstract**

Silicon-vacancy (SiV) centers in diamond are a promising candidate for all-optical nanoscale high-sensitivity thermometry because they have sufficient sensitivity to reach the subkelvin precision required for application to biosystems. It is expected that nanodiamonds with SiV centers can be injected into cells to measure the nanoscale local temperatures of biosystems such as organelles. However, the smallest particle size used to demonstrate thermometry using SiV centers is a few hundred nanometers. We recently developed SiV-center-containing nanodiamonds via a detonation process that is suitable for large-scale production. Here, we investigate the spectral response of SiV-center-containing detonation nanodiamonds (SiV-DNDs) to temperature. We used air-oxidized and polyglycerol-coated SiV-DNDs with a mean particle size of around 20 nm, which is the smallest size used to demonstrate thermometry using color centers in nanodiamond. We found that the zero-phonon line for SiV-DND is linearly red-shifted with increasing temperature in the range of 22.0 to 40.5 °C. The peak sensitivity to temperature was $0.011 \pm 0.002$ nm/K, which agrees with the reported high sensitivity of SiV centers in bulk diamond. A temperature sensitivity analysis revealed that SiV-DND thermometry can achieve subkelvin precision. All-optical SiV-DND thermometry will be important for investigating nanosystems such as organelles in living cells.

# 1. Introduction

The measurement of local temperature with nanoscale spatial resolution has a wide range of applications in science and technology. In particular, a nanoscale thermometer with subkelvin precision would be a powerful tool in the life sciences because temperature variation is common at the single-cell level due to biochemical reactions inside organelles.[1,2] For example, for measurement of the temperature inside the nucleus, particles with a size of a few tens of nanometers are required to penetrate a nuclear pore.[3]

Contactless fluorescent nanoscale thermometers, including those based on quantum dots, fluorescent dyes, fluorescent proteins, and nanodiamonds, have been extensively investigated.[4,5] Systems not based on nanodiamonds have absolute temperature uncertainties on the order of 1 K; however, due to their low spectral density and weak transition dipoles, these measurements typically require a long time to achieve subkelvin uncertainty.[6-9] Recently, the color centers in nanodiamonds have attracted attention for application to a high-sensitivity thermometer with a high spatial resolution.[10-15] These color centers have visible to near-infrared emissions. Temperature sensing based on the response of these emissions to temperature has been reported.[2,10-14] Nanodiamond has low cytotoxicity and can be surface-modified, making it suitable for modification to target cells and location tracking.[15] Among the color centers in nanodiamond, nitrogen-vacancy (NV) centers have promising potential for various biological applications from its optically detected magnetic resonance (ODMR) spectrum by utilizing visible and microwave excitation.[15] High-sensitivity temperature sensing based on the frequency shift of the ODMR spectrum has been demonstrated.[14] NV fluorescence imaging, temperature measurement, and temperature mapping of cells using contained nanodiamond have been reported.[2,16,17]

Silicon-vacancy (SiV) centers have also received interest as thermometers.[12,18-23] Although SiV centers have lower reported sensitivities than those of NV centers, they have several advantages. For example, they can be used to measure temperature optically without microwaves and their zero-phonon line (ZPL) width is very narrow. SiV centers have a sharp ZPL at 737 nm even at room temperature. The wavelengths of their excitation and emission are both in the window of biological samples (600-900 nm). Therefore, SiV centers are promising as a probe for bioimaging[19,20] and have attracted attention as a single-photon source.[21-23] Because the peak wavelength and intensity of the ZPL depend on temperature, all-optical temperature sensing has been demonstrated. High sensitivities of 0.36 K/$\sqrt{\text{Hz}}$ for an SiV ensemble in a bulk crystal and 0.54 K/$\sqrt{\text{Hz}}$ for SiV in nanodiamond with a particle size of 200 nm have been reported.[12]

The smallest reported particle size of SiV-center-containing nanodiamonds for temperature measurement is 200 nm.[12] Numerous methods for synthesizing nanodiamonds have been studied and developed, including chemical vapor deposition (CVD), high-pressure high-temperature (HPHT), and detonation processes.[24,25] The primary particles of detonation nanodiamonds (DNDs) are uniform, small (< 10 nm), and spherical and denotation has a high yield and low cost, which are very important for practical applications.[24-27] We recently reported the synthesis of SiV nanodiamonds with single-digit sizes via a straightforward detonation process.[28] Here, we investigate the temperature dependence of the emission spectrum of SiV-DNDs for nanoscale temperature sensing. We measured the temperature sensitivity using polyglycerol (PG)-coated SiV-DNDs with a mean particle size of about 20 nm. The smallest reported particle size of NV-center-containing nanodiamonds for temperature measurement is around 30 nm (achieved with HPHT nanodiamonds).[10] The recently reported temperature sensitivities of the color defect centers in nanodiamond are summarized in supplementary information. The proposed PG-coated SiV-DNDs thus have the smallest particle size of any color-center-containing nanodiamonds for temperature sensing.

## 2. Materials and Methods

For nanoscale all-optical temperature sensing, PG-coated SiV-DNDs were prepared according to previously reported procedures with some modifications.[28,29] A mixed explosive with silicon dopant (2,4,6-trinitrotoluene (TNT) : hexahydro-1,3,5-trinitro-1,3,5-triazine (RDX) : triphenylsilanol (TPS) = 59:40:1 wt%, 60 g for each detonation batch) was detonated under a $CO_2$ atmosphere. The detonation product contained agglomerated DNDs and numerous kinds of carbon soot and metal impurities. To remove non-diamond materials, the detonation product was purified with mixed acid ($H_2SO_4$ + $HNO_3$) at 150 °C for 5 h, added deionized (DI) water at 70 °C, and heated 150 °C again for 5 h. The collected sample was treated with 8 M NaOH at 70 °C for 8 h to remove the remaining acid and rinsed with DI water. After drying, the sample was air-oxidized in $O_2/N_2$ (4:96 vol%) at 570 °C for 8 h to remove $sp^2$ carbon on the DND's surface. The resulting powder was suspended in DI water at a concentration of 6.0 wt% and then sonicated with zirconia beads using an ultrasonic processor (Hielscher Ultrasound Technology, UP400S; electrical power: 400 W) at 50% amplitude for 180 min to obtain a well-dispersed water dispersion.

The dispersion was evaporated to dryness. The resulting dried powder (1.0 g) was suspended in ethylene glycol (1.0 g), and glycidol (5.0 g, 67.5 mmol) was added intermittently over 3 h with the temperature kept in the range of 105-110 °C. The resulting black dispersion was stirred at the same temperature for 5 h and subsequently at room temperature overnight. DI water was added to degrade the unreacted glycidol and the whole dispersion was diluted further with DI water to ~ 400 mL and condensed with an ultrafiltration membrane (Ultracel® membrane, 10 kDa) to < 10 mL. This process of dilution followed by condensation was repeated two times to obtain purified PG-coated SiV-DNDs. The PG-coating provides high dispersibility to the SiV-DNDs, enabling the centrifugation process easily. Finally, the sample was centrifuged at 30000 rpm (66,024*g*) for 12 min to remove large particles.

The particle size for the bare DNDs after air oxidization was characterized by powder XRD (Rigaku, SmartLab) with Cu-Kα1 radiation (λ = 1.54 Å). The average particle size was calculated by applying Scherrer's formula to the (111) diffraction peak. The particle size for the PG-coated DNDs after centrifugation was evaluated by dynamic light scattering (Microtrac Nanotrac II) in solution conditions. In addition, for the bare and PG-coated DNDs, TEM (JEOL, JEM-1400Plus; acceleration voltage: 120 kV) images were acquired of a few droplets on the water-suspended samples dried under ambient conditions on grids.

Figure 1(a) shows the structure of an SiV center in a diamond crystal. An SiV center consists of a divacancy in the crystal and a silicon atom located at the center of the divacancy. The inversion symmetric structure ($D_{3d}$) guarantees a stable and sharp ZPL at around 737 nm.[18] Figure 1(b) shows the chemical structure of the two explosives, TNT and RDX, and Si dopant, TPS, used in the detonation process. The detonation product was first purified by the processes of mixed acid, alkali, and air oxidization treatments in sequence. Next, the sample was dispersed by bead milling and then coated with PG. Finally, the sample was collected after centrifugation to remove large particles. Figure 1(c) shows a transmission electron microscopy (TEM) image of DNDs after the air oxidization process. Although the bare DNDs form strongly aggregated clusters, the size of each particle is around 10 nm. This result is supported by X-ray diffraction (XRD) analysis. As shown in Fig. 1(e), the XRD pattern has three peaks that correspond to diffraction from the (111), (220), and (311) planes of the diamond crystal structure. By applying Scherrer's formula to the (111) peak, the average particle size was calculated to be 10.8 nm. Figure 1(d) shows a TEM image of PG-coated DNDs after centrifugation. Because the particles were well separated, we directly evaluated the particle size distribution from the TEM image. The histogram of the particle size for 208 PG-coated DNDs peaks at around 20 nm (Fig. 1f). In addition, the results of dynamic light scattering in solution (Fig. 1g) show that the particle size peaks at 12 nm, which basically agrees with the result from the TEM image. Although the particle size is larger than that for bare DNDs due to the PG coating, it is sufficiently small for biological applications inside a cell. Hereafter, we evaluate only PG-coated DNDs after centrifugation.

## 3.  Results and Discussion

We first confirmed the spectral properties of SiV-DNDs using a home-built confocal microscope. Figure 2(a) shows a schematic diagram of the optical setup. For details, see the supplementary information. Briefly, an SiV-DND suspension was dropped onto a glass coverslip and dried in an air atmosphere. A 532-nm laser beam was focused from the backside by an oil-immersion objective lens, and the photoluminescence (PL) of the sample was collected by the same objective. The PL signal was introduced to either a single-photon counting module for the PL image or a monochromator for the PL spectrum.

Figure 2(b) shows a PL image of SiV-DNDs. Owing to the high concentration of the sample, the whole region has a stable and high PL intensity. In addition, bright spots with stronger PL intensities appear. Figure 2(c) shows PL spectra of a typical bright spot A in Fig. 2(b), and a typical region without spots, region B in Fig. 2(b). The spectrum of the spot A consists of a sharp spectral component at around 737 nm and a broad spectral component ranging from 550 to 900 nm. Although the spectrum of region B also has sharp and broad spectral components, the intensity of the sharp component is much smaller than that for the spot A. As shown in Fig. 2(d), the difference spectrum between A and B clearly shows the typical spectrum of an SiV center.[23] The broad spectrum is considered to consist of DND-originated emissions, such as those associated with residual inner amorphous or $sp^2$ carbon, other possible color centers,[30] or silicon-related defects with an unknown structure.[31] Note that the negatively charged NV center (NV⁻) may not be a major component of the broad spectrum because there was no ODMR signal at 2.87 GHz in the absence of a static magnetic field in another experiment (data not shown).[15]

By taking advantage of the sharpness of the ZPL for the SiV centers, the background emission (i.e., emission other than that due to SiV centers) can be well eliminated by conducting the measurement with a narrow bandpass filter, as shown in Fig. 3(a). Figures 3(b) and 3(c) show images measured with and without the filter, respectively. As shown, the background emission is well eliminated when the filter is used. The contrast between the SiV emission and the background emission increases almost three fold (supplementary information). If SiV-DNDs are applied in a biological system and a narrow bandpass filter that passes only the ZPL emission is used, we can eliminate significant parts of the autofluorescence of the system, making it much easier to identify the emission from the SiV-DNDs.

We next investigated the change in the SiV-DND spectrum with temperature. For this measurement, we added a temperature control system to the microscope (supplementary information). As shown in Fig. 2(a), the sample coverslip was placed between a thermoplate and a copper plate with tight contact to realize quick thermal equilibrium. The sample temperature was controlled by changing the preset temperatures of the thermoplate and a lens heater attached to the objective and monitored via a thermistor embedded in the copper plate.

Figure 4(a) shows a high-resolution PL spectrum of a typical bright spot measured at sample temperatures ($T$) of 22.0 and 40.5 °C. The ZPL peak wavelength ($\lambda_p$) shifted to red and the linewidth broadened with increasing the temperature (about PL spectrum analysis, see supplementary information). Both the peak shift and linewidth broadening are mainly due to the second-order electron-phonon interaction, especially elastic Raman scattering for the SiV center. [32] In the current study, we focused on the ZPL wavelength peak shift. A detailed study revealed that the SiV center's optical transition energy between the electron-ground and electron-excited states reduces proportionally to $T^2$ ranging from 298 K up to 873 K (25 °C - 600 °C).[33] The ZPL peak wavelength also shifts to red according to the temperature dependence of this transition energy. However, we can assume $\lambda_p$ linearly shifts with $T$ for the range of interest, which corresponds to thermometry in biological systems, as done in previous research.[11,12] As shown in Fig. 4(b), the experimental data can be well fitted by a linear function with a ZPL peak sensitivity to temperature ($\Delta\lambda/\Delta T$) of 0.015 nm/K. This result agrees with those for an SiV center ensemble in bulk diamond ($\Delta\lambda/\Delta T = 0.012$ nm/K,[11] 0.0124 nm/K,[12]) and SiV-containing HPHT nanodiamonds (0.0129 nm/K[34]). For thermometry in biological systems using SiV-DNDs, it is important to estimate the temperature sensitivity $\eta_T$. In this research, we define $\eta_T$ as the temperature uncertainty $\sigma_T$ for a unit integration time ($t_{int} = 1$ s), as done in previous work.[12] Figure 4(c) shows $\sigma_T$ for a bright spot as a function of $t_{int}$ measured at $T = 22$ °C. Here, $\sigma_T$ can be converted using the ZPL peak position uncertainty $\sigma_\lambda$ (the standard deviation of the fitting parameter $\lambda_p$) and the slope $\Delta\lambda/\Delta T$, that is, $\sigma_T = \sigma_\lambda / (\Delta\lambda/\Delta T)$. For $t_{int}$ below 10 s, $\sigma_T$ rapidly decreases according to a curve proportional to $1/\sqrt{t_{int}}$. Because the detected photon number $N_{int}$ is proportional to $t_{int}$, the photon shot noise has a dominant effect on sensitivity. As shown in Fig. 4(c), we estimated $\eta_T = 1.1$ K/$\sqrt{Hz}$ at $t_{int} = 1$ s. For $t_{int}$ above 10 s, $\sigma_T$ keeps decreasing but gradually deviates from the curve of $1/\sqrt{t_{int}}$. This is because other factors, such as the drift of the sample position and the stability of the whole microscope system, also affect the temperature uncertainty. Despite this deviation, $\sigma_T$ reached 0.4 K for an integration time of 30 s. Nanodiamonds with subkelvin temperature precision have potential for realizing all-optical living cell thermometry.

In addition to the above bright spot, we measured the PL spectral response to temperature for more than 30 bright spots randomly selected from a confocal image. For all the bright spots, $\lambda_p$ was red-shifted with increasing temperature. 80% of the measured spots had good linearity between $\lambda_p$ and temperature. Figure 4(d) summarizes the distribution between $\Delta\lambda/\Delta T$ for 26 bright spots. The average value of the slope $\Delta\lambda/\Delta T$ was $0.011 \pm 0.002$ nm/K, which agrees with previous measurements of an SiV ensemble in bulk diamond.[12] The scattering of the slopes in DNDs, which is shown in Fig. 4(d), was larger than that in bulk diamond.[12] This may be due to the small particle size of DNDs and differences in the surrounding environment.

The average value of $\eta_T$ is $2.9 \pm 1.3$ K/$\sqrt{\text{Hz}}$, which indicates that subkelvin precision can be realized for an integration time of 10 s using the measured spots. Table 1 summarizes all the parameters, including $\lambda_p$ and the spectral width (full width at half maximum, FWHM) $2\gamma$, for the 26 spots. For the histograms of $\lambda_p$, $2\gamma$, and $\eta_T$, see supplementary information. It should be noted that $\eta_T$ depends not only on $\Delta\lambda/\Delta T$ but also on the number of SiV centers in the measured spot. The $\eta_T$ is distributed widely compared to $\Delta\lambda/\Delta T$, which is considered to be due to the difference in the number of SiV centers among the measured spots and/or differences in the surrounding environment of each bright spot. Next, we discuss the number of SiV centers and the potential temperature sensitivity of individual nanodiamonds.

The signal intensities of the bright spots were much larger than that of a single SiV center, which can be deduced from that of a single NV center.[22] In our sample, the number of the SiV centers included in each nanodiamond is not estimated. In general, DNDs tend to form aggregates via Coulomb interaction.[35] Large SiV-DNDs particles and/or aggregates were removed by dispersion treatment with a bead milling and ultracentrifugation after PG coating. However, the SiV-DNDs probably reaggregated despite the PG coating because the SiV-DNDs suspension was dropped and dried onto the cover glass in our measurement.

As shown in Fig. 4(c), the measurement uncertainty for the bright spot follows the shot-noise limit $1/\sqrt{t_{\text{int}}}$, suggesting that the precision and sensitivity can be improved by increasing the SiV density or photon collection rate. Under the shot-noise limit, the temperature sensitivity $\eta_T$ depends on the number of SiV centers ($N$) as $\eta_T \propto 1/\sqrt{N}$. Therefore, the sensitivity of a single SiV center in a DND can be discussed from the results. The sensitivity of a single SiV center is an essential figure of merit toward the use of a single or a small number of nanodiamonds to obtain nanoscale spatial resolution. A previous study[12] under the shot-noise limit reported temperature sensitivities of 360 mK/$\sqrt{\text{Hz}}$ with about 100 SiV centers in bulk and 521 mK/$\sqrt{\text{Hz}}$ with fewer than 10 SiV centers in nanodiamond (diameter: 200 nm). The number of SiV centers in the bright spots, where the sensitivity is $\eta_T = 1.1$ K/$\sqrt{\text{Hz}}$, is estimated to be about 60 SiV centers based on a previous report in which a single NV and a single SiV were measured and compared in terms of focal volume.[22] The details of the estimation are described in the supplementary information. From the estimated number of SiV centers and $\eta_T = 1.1$ K/$\sqrt{\text{Hz}}$, the precision for a single SiV center is estimated to approach 1 K in a measurement time of about 73 s under the shot-noise limit. Therefore, if multiple SiV centers can be introduced into a single nanodiamond, the temperature precision would reach the subkelvin level in a measurement time shorter than one minute using only a single nanodiamond. This indicates that SiV-DNDs have potential as all-optical nanoscale temperature sensors with subkelvin precision.

## 4. Conclusions

In summary, we investigated the optical temperature dependence of SiV-DNDs. To the best of our knowledge, PG-coated SiV-DNDs have the smallest size for thermometry of all color-center-containing nanodiamonds. This small size allows SiV-DNDs to be introduced into a living cell without invasive or complicated methods. The ZPL peak wavelength for the aggregated SiV-DNDs was linearly red-shifted with increasing temperature. The wavelength sensitivity to temperature was comparable to those for an SiV ensemble in bulk diamond and HPHT SiV-NDs. In addition, thermometry based on SiV-DNDs can achieve subkelvin precision.

Although we focused to evaluate the temperature sensitivity based on the ZPL peak wavelength shift, several parameters such as the ZPL spectral width are also possible candidates to perform all-optical thermometry.[11,12] In addition, Choi et al. conducted a multiparametric analysis using the peak wavelength, spectral width, and amplitude for 250-nm diamond nanocrystals and found that the intrinsic noise floor is about 13 mK/$\sqrt{\text{Hz}}$.[36] This noise level is equivalent to a 1000-fold decrease of the integration time compared with that of conventional all-optical methods.[36] Multiparametric analysis would thus improve the sensitivity of our temperature sensor. In addition, increasing the concentration of SiV centers in individual nanodiamonds is essential for high-sensitivity measurements. These developments would allow the measurement of nanoscale local temperature of organelles in a living cell.

**Acknowledgments**

The authors are grateful for financial support from the Japanese Ministry of Education, Culture, Sports, Science and Technology (MEXT)-QLEAP project (PMXS0120330644) and a Kakenhi Grant-in-Aid (No. 21H04653) from the Japan Society for the Promotion of Science (JSPS).

**References**

1    Bai, T.; Gu, N. Micro/Nanoscale Thermometry for Cellular Thermal Sensing. *Small* **2016**, *12*, 4590–4610.

2    Kucsko, G.; Maurer, P. C.; Yao, N. Y.; Kubo, M.; Noh, H. J.; Lo, P. K.; Park, H.; Lukin, M. D. Nanometre-Scale Thermometry in a Living Cell. *Nature* **2013**, *500*, 54–58.

3    Pan, L.; Liu, J.; Shi, J.; Cancer cell nucleus-targeting nanocomposites for advanced tumor therapeutics. *Chem. Soc. Rev.*, **2018**, *47*, 6930-6946.

4     Bradac, C.; Lim, S. F.; Chang, H. C.; Aharonovich, I. Optical Nanoscale Thermometry: From Fundamental Mechanisms to Emerging Practical Applications. *Adv. Opt. Mater.* **2020**, *8*, 1–29.

5     Zhou, J.; del Rosal, B.; Jaque, D.; Uchiyama, S.; Jin, D. Advances and Challenges for Fluorescence Nanothermometry. *Nat. Methods* **2020**, *17*, 967–980.

6     Liu, H.; Fan, Y.; Wang, J.; Song, Z.; Shi, H.; Han, R.; Sha, Y.; Jiang, Y. Intracellular Temperature Sensing: An Ultra-Bright Luminescent Nanothermometer with Non-Sensitivity to PH and Ionic Strength. *Sci. Rep.* **2015**, 5, 14879.

7     Vetrone, F.; Naccache, R.; Zamarrón, A.; De La Fuente, A. J.; Sanz-Rodríguez, F.; Maestro, L. M.; Rodriguez, E. M.; Jaque, D.; Sole, J. G.; Capobianco, J. A. Temperature Sensing Using Fluorescent Nanothermometers. *ACS Nano* **2010**, 4, 3254–3258.

8     Kalytchuk, S.; Poláková, K.; Wang, Y.; Froning, J. P.; Cepe, K.; Rogach, A. L.; Zbořil, R. Carbon Dot Nanothermometry: Intracellular Photoluminescence Lifetime Thermal Sensing. *ACS Nano* **2017**, 11, 1432–1442.

9     Yue, Y.; Wang, X. Nanoscale Thermal Probing. *Nano Rev.* **2012**, 3, 11586.

10    Plakhotnik, T.; Doherty, M. W.; Cole, J. H.; Chapman, R.; Manson, N. B. All-Optical Thermometry and Thermal Properties of the Optically Detected Spin Resonances of the NV- Center in Nanodiamond. *Nano Lett.* **2014**, *14*, 4989–4996.

11    Alkahtani, M.; Cojocaru, I.; Liu, X.; Herzig, T.; Meijer, J.; Küpper, J.; Lühmann, T.; Akimov, A. V.; Hemmer, P. R. Tin-Vacancy in Diamonds for Luminescent Thermometry. *Appl. Phys. Lett.* **2018**, *112*, 241902.

12    Nguyen, C. T.; Evans, R. E.; Sipahigil, A.; Bhaskar, M. K.; Sukachev, D. D.; Agafonov, V. N.; Davydov, V. A.; Kulikova, L. F.; Jelezko, F.; Lukin, M. D. All-Optical Nanoscale Thermometry with Silicon-Vacancy Centers in Diamond. *Appl. Phys. Lett.* **2018**, *112*, 203102.

13   Fan, J. W.; Cojocaru, I.; Becker, J.; Fedotov, I. V.; Alkahtani, M. H. A.; Alajlan, A.; Blakley, S.; Rezaee, M.; Lyamkina, A.; Palyanov, Y. N.; Borzdov, Y. M.; Yang, Y. P.; Zheltikov, A.; Hemmer, P.; Akimov, A. V. Germanium-Vacancy Color Center in Diamond as a Temperature Sensor. *ACS Photonics* **2018**, *5*, 765–770.

14   Hayashi, K.; Matsuzaki, Y.; Taniguchi, T.; Shimo-Oka, T.; Nakamura, I.; Onoda, S.; Ohshima, T.; Morishita, H.; Fujiwara, M.; Saito, S.; Mizuochi, N. Optimization of temperature sensitivity using the optically detected magnetic-resonance spectrum of a nitrogen-vacancy center ensemble. *Phys. Rev. Appl.*, **2018**, 10, 034009.

15   Schirhagl, R.; Chang, K.; Loretz, M.; Degen, C. L. Nitrogen-Vacancy Centers in Diamond: Nanoscale Sensors for Physics and Biology. *Annu. Rev. Phys. Chem.* **2014**, *65*, 83–105.

16   Fujiwara, M.; Sun, S.; Dohms, A.; Nishimura, Y.; Suto, K.; Takezawa, Y.; Oshimi, K.; Zhao, L.; Sadzak, N.; Umehara, Y.; Teki, Y.; Komatsu, N.; Benson, O.; Shikano, Y.; Kage-Nakadai, E. Real-Time Nanodiamond Thermometry Probing in-Vivo Thermogenic Responses. *Sci. Adv.* **2020**, *6*, No. eaba9636.

17   Simpson, D. A.; Morrisroe, E.; McCoey, J. M.; Lombard, A. H.; Mendis, D. C.; Treussart, F.; Hall, L. T.; Petrou, S.; Hollenberg, L. C. L. Non-Neurotoxic Nanodiamond Probes for Intraneuronal Temperature Mapping. *ACS Nano* **2017**, *11*, 12077–12086.

18   Bradac, C.; Gao, W.; Aharonovich, I.; Forneris, J.; Trusheim, M. E. Quantum Nanophotonics with Group IV Defects in Diamond. *Nat. Commun.* **2019**, *10*, 5625.

19   Alkahtani, M. H.; Alghannam, F.; Jiang, L.; Rampersaud, A. A.; Brick, R.; Gomes, C. L.; Scully, M. O.; Hemmer, P. R. Fluorescent Nanodiamonds for Luminescent Thermometry in the Biological Transparency Window. *Opt. Lett.* **2018**, *43*, 3317.

20    Merson, T. D.; Castelletto, S.; Aharonovich, I.; Turbic, A.; Kilpatrick, T. J.; Turnley, A. M. Nanodiamonds with silicon vacancy defects for nontoxic photostable fluorescent labeling of neural precursor cells. *Opt. Lett.* **2013**, 37, 4170-4173.

21    Neu, E.; Steinmetz, D.; Riedrich-Möller, J.; Gsell, S.; Fischer, M.; Schreck, M.; Becher, C. Single Photon Emission from Silicon-Vacancy Colour Centres in Chemical Vapour Deposition Nano-Diamonds on Iridium. *New J. Phys.* **2011**, *13*, 025012.

22    Rogers, L. J.; Jahnke, K. D.; Teraji, T.; Marseglia, L.; Müller, C.; Naydenov, B.; Schauffert, H.; Kranz, C.; Isoya, J.; McGuinness, L. P.; Jelezko, F. Multiple Intrinsically Identical Single-Photon Emitters in the Solid State. *Nat. Commun.* **2014**, *5*, 1–6.

23    Berhane, A. M.; Choi, S.; Kato, H.; Makino, T.; Mizuochi, N.; Yamasaki, S.; Aharonovich, I. Electrical excitation of silicon-vacancy centers in single crystal diamond, *Appl. Phys Lett.*, **2015**, *106*, 171102.

24    J.-C. Arnault, Nanodiamonds: advanced material analysis, properties and applications, 1st ed., Elsevier, Amsterdam, 2017, pp. 166–172, pp. xix, 25–31

25    V.N. Mochalin, O. Shenderova, D. Ho, Y. Gogotsi, The properties and applications of nanodiamond, *Nat. Nanotechnol.* **2012**, *7*, 11–23.

26    Sotoma, S.; Terada, D.; Segawa, T. F.; Igarashi, R.; Harada, Y.; Shirakawa, M. Enrichment of ODMR-Active Nitrogen-Vacancy Centres in Five-Nanometre-Sized Detonation-Synthesized Nanodiamonds: Nanoprobes for Temperature, Angle and Position. *Sci. Rep.* **2018**, *8*, 1–8.

27    Terada, D.; Segawa, T. F.; Shames, A. I.; Onoda, S.; Ohshima, T.; Osawa, E.; Igarashi, R.; Shirakawa, M. Monodisperse Five-Nanometer-Sized Detonation Nanodiamonds Enriched in Nitrogen-Vacancy Centers. *ACS Nano* **2019**, *13*, 6461–6468.

28    Makino, Y.; Mahiko, T.; Liu, M.; Tsurui, A.; Yoshikawa, T.; Nagamachi, S.; Tanaka, S.; Hokamoto, K.; Ashida, M.; Fujiwara, M.; Mizuochi, N.; Nishikawa, M. Straightforward

Synthesis of Silicon Vacancy (SiV) Center-Containing Single-Digit Nanometer Nanodiamonds via Detonation Process. *Diam. Relat. Mater.* **2021**, *112*, 108248.

29    Yoshikawa, T.; Liu, M.; Chang, S. L. Y.; Kuschnerus, I. C.; Makino, Y.; Tsurui, A.; Mahiko, T.; Nishikawa, M. Steric interaction of polyglycerol-functionalized detonation nanodiamonds, *Langmuir*, **2022**, *38*, 661–669.

30    Neu, E.; Arend, C.; Gross, E.; Guldner, F.; Hepp, C.; Steinmetz, D.; Zscherpel, E.; Ghodbane, S.; Sternschulte, H.; Steinmüller-Nethl, D.; et al. Narrowband Fluorescent Nanodiamonds Produced From Chemical Vapor Deposition Films. *Appl. Phys. Lett.* **2011**, *98*, 243107.

31    Lindner, S.; Bommer, A.; Muzha, A.; Krueger, A.; Gines, L.; Mandal, S.; Williams, O.; Londero, E.; Gali, A.; Becher, C. Strongly Inhomogeneous Distribution of Spectral Properties of Silicon-Vacancy Color Centers in Nanodiamonds. *New J. Phys.* **2018**, *20*, 115002.

32    Jahnke, K. D.; Sipahigil, A.; Binder, J. M.; Doherty, M. W.; Metsch, M.; Rogers, L. J.; Manson, N. B.; Lukin, M. D.; Jelezko, F. Electron-Phonon Processes of the Silicon-Vacancy Centre in Diamond. *New Journal of Physics* **2015**, *17*, 043011.

33    Zaghrioui, M.; Agafonov, V. N.; Davydov, V. A. Nitrogen and Group-IV (Si, Ge) Vacancy Color Centres in Nano-Diamonds: Photoluminescence Study at High Temperature (25 °c-600 °c). *Mater. Res. Express* **2020**, *7*, 015043.

34    Choi, S.; Agafonov, V. N.; Davydov, V. A.; Kulikova, L. F.; Plakhotnik, T. Formation of Interstitial Silicon Defects in Si- And Si,P-Doped Nanodiamonds and Thermal Susceptibilities of SiV- Photoluminescence Band. *Nanotechnology* **2020**, *31*, 205079.

35    Chang, S. L. Y.; Reineck, P.; Williams, D.; Bryant, G.; Opletal, G.; El-Demrdash, S. A.; Chiu, P. L.; Osawa, E.; Barnard, A. S.; Dwyer, C. Dynamic Self-Assembly of Detonation Nanodiamond in Water. *Nanoscale* **2020**, *12*, 5363–5367.

36    Choi, S.; Agafonov, V. N.; Davydov, V. A.; Plakhotnik, T. Ultrasensitive All-Optical Thermometry Using Nanodiamonds with a High Concentration of Silicon-Vacancy Centers and Multiparametric Data Analysis. *ACS Photonics* **2019**, *6*, 1387–1392.

**Table and Table captions**

**Table 1.** Summary of parameters for 26 bright spots. The ZPL peak wavelength ($\lambda_p$) and the spectral width (FWHM) $2\gamma$ are shown as average values. For $\Delta\lambda/\Delta T$, $\eta_T$, and $\sigma_T$ at $t_{int}$ = 10 s, both the average and the best values are displayed. All parameters except $\Delta\lambda/\Delta T$ were obtained at a sample temperature of 22.0 °C.

| data point | $\lambda_{p,\,avg}$ [nm] | $2\gamma_{avg}$ [nm] | $(\Delta\lambda/\Delta T)_{avg}$ [nm/K] | $(\Delta\lambda/\Delta T)_{best}$ [nm/K] | $\eta_{T,\,avg}$ [K/$\sqrt{\text{Hz}}$] | $\eta_{T,\,best}$ [K/$\sqrt{\text{Hz}}$] | $\sigma_{T,\,avg}$ ($t_{int}$ = 10 s) [K] | $\sigma_{T,\,best}$ ($t_{int}$ = 10 s) [K] |
|---|---|---|---|---|---|---|---|---|
| 26 | 737.09 ± 0.22 | 7.09 ± 0.95 | 0.011 ± 0.002 | 0.016 | 2.9 ± 1.3 | 1.1 | 1.0 ± 0.4 | 0.5 |

**Figure Captions**

**Figure 1.** SiV-center-containing detonation nanodiamonds (SiV-DNDs). (a) SiV center in diamond crystal lattice. The silicon atom, split vacancy site, and surrounding carbon atoms are labeled. (b) Chemical structure of two explosives, TND and RDX, and Si dopant, TPS, used in detonation process. TEM images of (c) bare SiV-DNDs after air oxidation and (d) polyglycerol (PG)-coated SiV-DNDs after centrifugation. The scale bar in each image is 100 nm. (e) X-ray diffraction pattern for bare SiV-DNDs. Three peaks correspond to diffractions from (111), (220), and (311) planes of the diamond crystal structure, respectively. (f) Particle size histogram of PG-coated SiV-DNDs analyzed from TEM image. (f) Size distribution of PG-coated SiV-DNDs in solution measured by dynamic light scattering.

**Figure 2**. (a) Optical setup for determining thermal shift of ZPL peaks of SiV-DNDs. (b) Confocal image of SiV-DNDs on coverslip. (c) Photoluminescence (PL) spectra of bright spot A and B (without bright spots) in (b) (white dotted circles). (d) Difference spectrum from spot A to region B. These measurements were performed at a temperature of 22.0 °C and the spectra

were measured using a monochromator with a 150-gr/mm diffraction grating (0.6-nm spectral resolution). The integration time for the measurement ($t_{int}$) was 10 s for each spectrum.

**Figure 3**. (a) Transmission curve for a 747 ± 20 nm bandpass filter. A PL spectrum of a typical bright spot is also shown. Confocal images of SiV-DNDs measured (b) with and (c) without bandpass filter.

**Figure 4**. (a) Typical PL spectra of SiV-DNDs at temperatures of 22.0 °C (blue curve) and 40.5 °C (red curve). The ZPL peak is red-shifted at higher temperatures. The integration time $t_{int}$ was 150 s for each spectrum. (b) ZPL peak wavelength ($\lambda_p$) as function of sample temperature. The peak sensitivity to temperature ($\Delta\lambda/\Delta T$) was obtained by a line fit (red line) to the experimental data (blue dots). (c) Temperature uncertainty ($\sigma_T$) as function of $t_{int}$ measured at temperature of 22.0 °C. The blue dots show experimental data and the red line indicates a fitting curve proportional to $1/\sqrt{t_{int}}$. (d) Histogram of $\Delta\lambda/\Delta T$ for 26 measured bright spots. The average value is 0.011 ± 0.002 nm/K. In this figure, we analyzed the spectra measured by a monochromator with an 1800-gr/mm diffraction grating (0.02-nm spectral resolution).

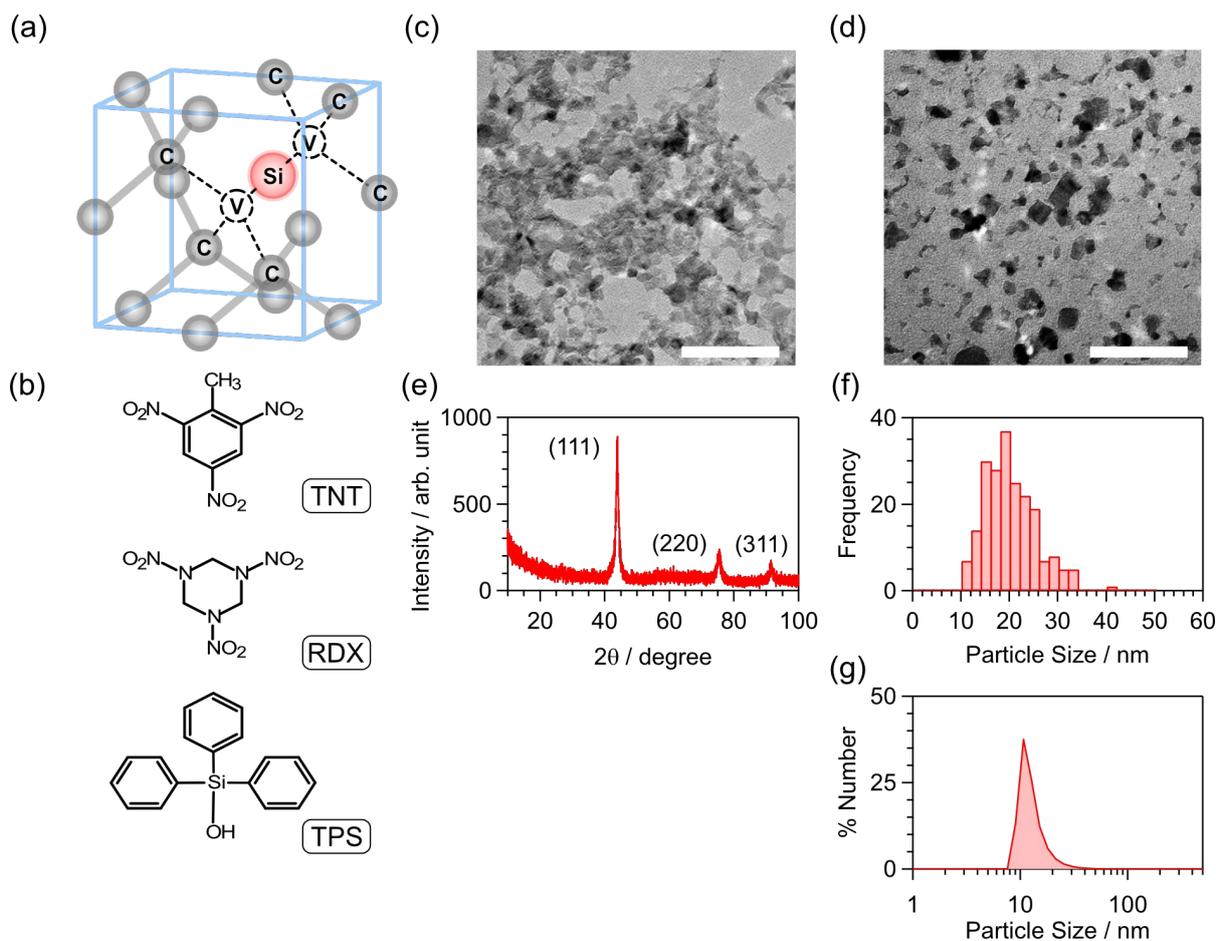

**Figure 1.** SiV-center-containing detonation nanodiamonds (SiV-DNDs). (a) SiV center in diamond crystal lattice. The silicon atom, split vacancy site, and surrounding carbon atoms are labeled. (b) Chemical structure of two explosives, TND and RDX, and Si dopant, TPS, used in detonation process. TEM images of (c) bare SiV-DNDs after air oxidation and (d) polyglycerol (PG)-coated SiV-DNDs after centrifugation. The scale bar in each image is 100 nm. (e) X-ray diffraction pattern for bare SiV-DNDs. Three peaks correspond to diffractions from (111), (220), and (311) planes of the diamond crystal structure, respectively. (f) Particle size histogram of PG-coated SiV-DNDs analyzed from TEM image. (f) Size distribution of PG-coated SiV-DNDs in solution measured by dynamic light scattering.

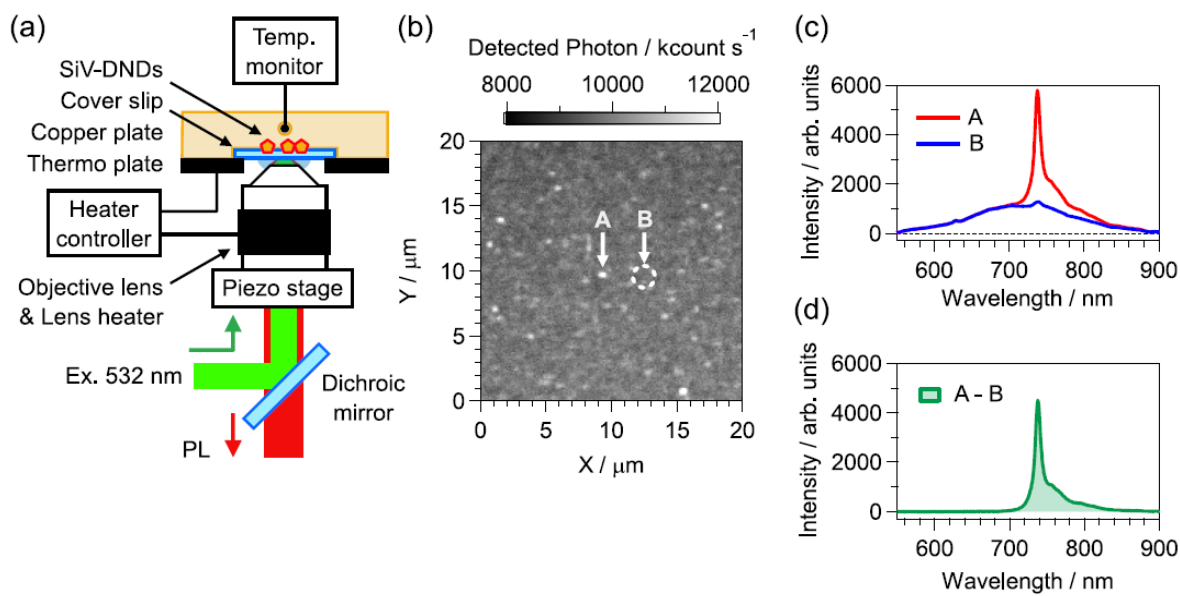

**Figure 2**. (a) Optical setup for determining thermal shift of ZPL peaks of SiV-DNDs. (b) Confocal image of SiV-DNDs on coverslip. (c) Photoluminescence (PL) spectra of bright spot A and B (without bright spots) in (b) (white dotted circles). (d) Difference spectrum from spot A to region B. These measurements were performed at a temperature of 22.0 °C and the spectra were measured using a monochromator with a 150-gr/mm diffraction grating (0.6-nm spectral resolution). The integration time for the measurement ($t_{int}$) was 10 s for each spectrum.

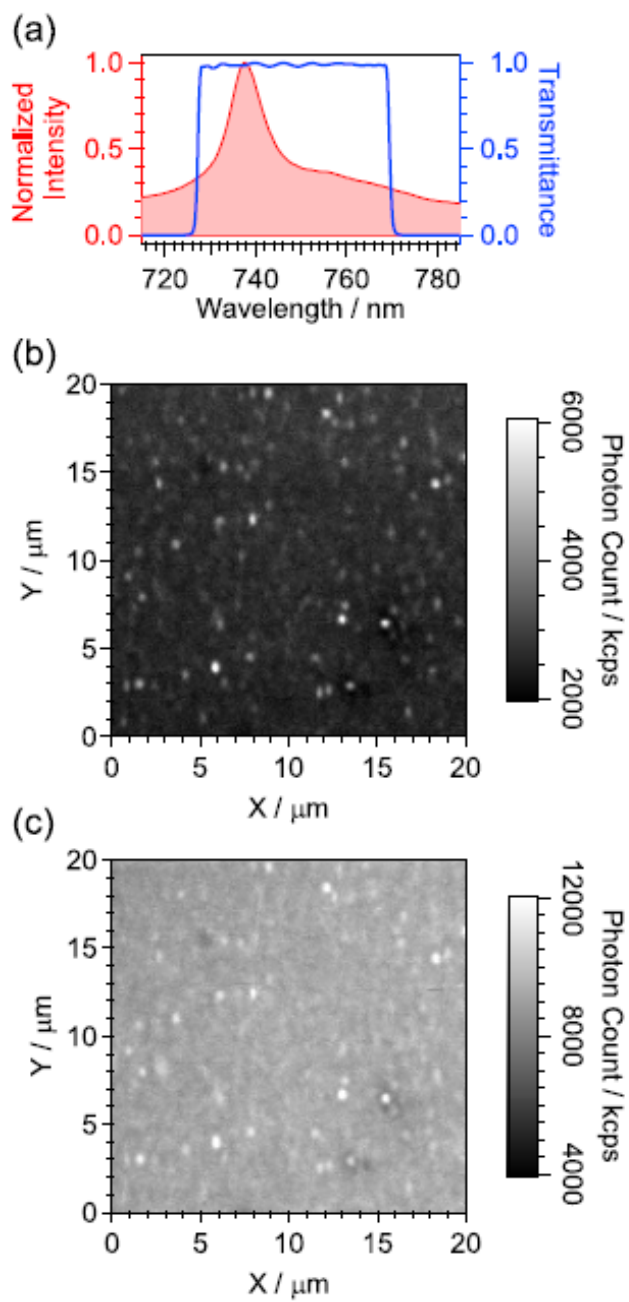

**Figure 3**. (a) Transmission curve for a 747 ± 20 nm bandpass filter. A PL spectrum of a typical bright spot is also shown. Confocal images of SiV-DNDs measured (b) with and (c) without bandpass filter.

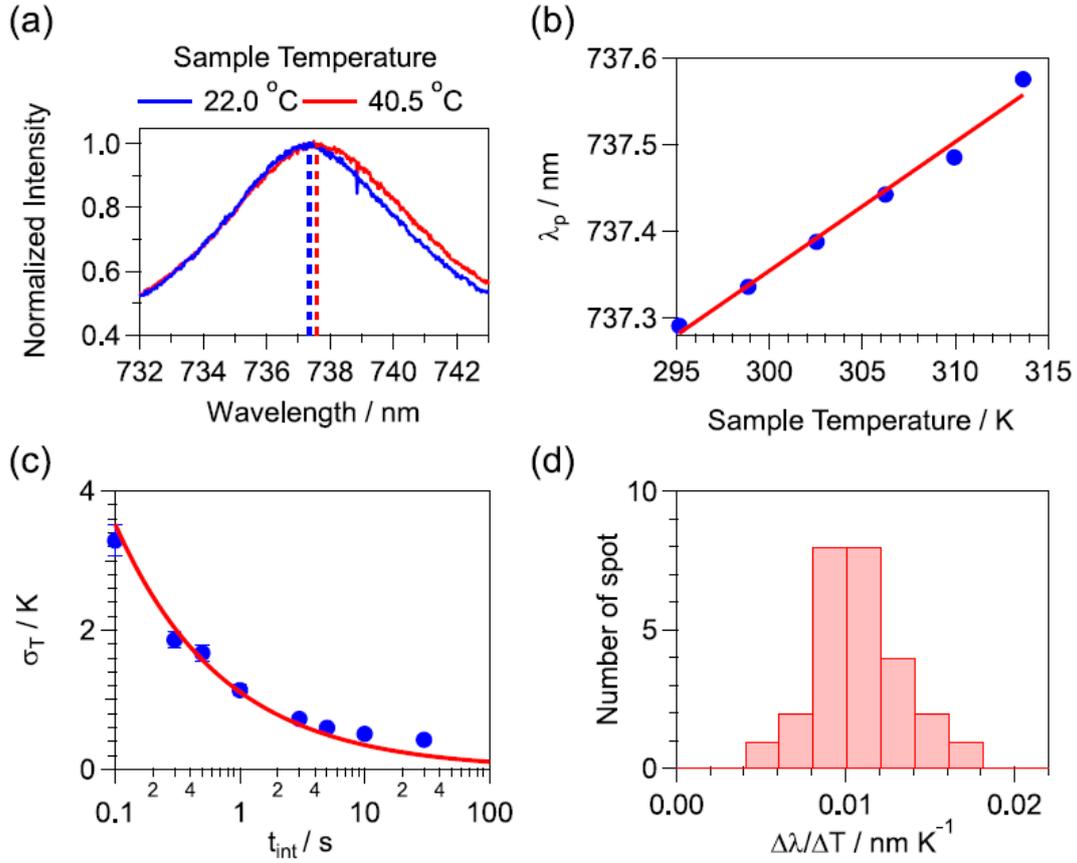

**Figure 4**. (a) Typical PL spectra of SiV-DNDs at temperatures of 22.0 °C (blue curve) and 40.5 °C (red curve). The ZPL peak is red-shifted at higher temperatures. The integration time $t_{int}$ was 150 s for each spectrum. (b) ZPL peak wavelength ($\lambda_p$) as function of sample temperature. The peak sensitivity to temperature ($\Delta\lambda/\Delta T$) was obtained by a line fit (red line) to the experimental data (blue dots). (c) Temperature uncertainty ($\sigma_T$) as function of $t_{int}$ measured at temperature of 22.0 °C. The blue dots show experimental data and the red line indicates a fitting curve proportional to $1/\sqrt{t_{int}}$. (d) Histogram of $\Delta\lambda/\Delta T$ for 26 measured bright spots. The average value is 0.011 ± 0.002 nm/K. In this figure, we analyzed the spectra measured by a monochromator with an 1800-gr/mm diffraction grating (0.02-nm spectral resolution).

**All-optical nanoscale thermometry based on silicon-vacancy centers in detonation nanodiamonds**

Masanori Fujiwara[1], Gaku Uchida[1], Izuru Ohki[1], Ming Liu[2], Akihiko Tsurui[2], Taro Yoshikawa[2], Masahiro Nishikawa[2], and Norikazu Mizuochi[1,3]

[1] Institute for Chemical Research, Kyoto University, Gokasho, Uji, Kyoto 611-0011, Japan

[2] Innovation and Business Development Headquarters, Daicel Corporation, 1239, Shinzaike, Aboshi-Ku, Himeji, Hyogo 671-1283, Japan

[3] Center for Spintronics Research Network, Kyoto University, Gokasho, Uji, Kyoto 611-0011, Japan

**Optical setup.**

The photoluminescence (PL) images of SiV-center-containing detonation nanodiamonds (SiV-DNDs) were acquired using a home-built confocal microscope. A 532-nm laser beam was focused by an oil-immersion objective lens (Lambda 100X Oil, Nikon; numerical aperture = 1.45) from the backside of the coverslip. The PL signal from the sample on the coverslip was collected by the same objective, filtered by a 552-nm dichroic mirror (LM01-552, Semrock), and passed through a 50-μm pinhole. Then, after being filtered by a 645-nm long-pass filter (BLP01-633R, Semrock), the signal was finally detected by a single-photon counting module (SPCM-AQRH014, Excelitas). To obtain the confocal image, the objective lens was placed on a three-axis piezo stage (NPXY100Z25-219, nPoint), which controlled the lens position. After 2D raster scanning of the objective at the optimal axial position, we found the bright spots in the image. The laser power was set to 1 mW before the objective lens. The lateral ($r_{\mathrm{FWHM}}$) and axial ($z_{\text{FWHM}}$) resolutions of the microscope were 0.4 and 1.5 μm, respectively, which were evaluated from the FWHM of the PL image of a single NV center in bulk diamond.

For the PL spectrum measurement, the PL signal was introduced to a Czerny-Turner monochromator (Acton SP2300i and SPEC-10, Princeton Instruments) by a mirror on a 90° flip mount. This mount was placed between the pinhole stage and the long-pass filter mount. We used two diffraction gratings in the monochromator, namely a 150 gr/mm grating (~350-nm spectral range and 0.6-nm spectral resolution) and an 1800 gr/mm grating (~20-nm range and 0.02-nm resolution). The former was used for the measurements over a wide wavelength range, for example Fig. 2(c), and the latter was used for the measurements with a high spectral resolution, for example Fig. 4(a). For the latter case, the 50-μm pinhole was replaced with a 100-μm pinhole to detect more PL signals.

**Contrast between SiV emission and background emission.**

To obtain a confocal image without a background signal Fig. 3(b), we used a 747 ± 20 nm bandpass filter (FF01-747/33, Semrock). The filter was fixed on another flip mount placed before the 645-nm long-pass filter mount. The transmission window corresponds to the wavelength region of the ZPL and the phonon sideband of the SiV center.

We evaluated the contrast between the SiV emission and the background emission as follows. First, we selected the diffraction image of a bright spot from a confocal image. Next, we fitted the diffraction pattern by the following 2D Gaussian function:

$$G(x, y) = A \exp\left[-\frac{1}{2}\left(\left(\frac{x - x_0}{w_x}\right)^2 + \left(\frac{y - y_0}{w_y}\right)^2\right)\right] + B,$$

where $A$ corresponds to the SiV center emission, $B$ is the background emission, $x_0$ ($y_0$) is the $x$ ($y$) coordinate of the peak position, and $w_x$ ($w_y$) is the width in the $x$ ($y$) direction. Finally, we obtained the contrast $C$ as $C = A/B$. We selected a total of 16 bright spots and calculated the contrast *C* for each one. The average contrast was 1.36 ± 0.40 and 0.46 ± 0.12 for images with and without the bandpass filter, respectively. From these results, we calculated the ratio of these contrast values as (1.36 ± 0.40)/(0.46 ± 0.12) = 3.0 ± 1.2. This indicates that the contrast was three-fold higher when using the bandpass filter.

**Temperature control.**

The sample temperature was controlled by a thermoplate and a lens heater (TPi-100RH26 and TPiE-LH, Tokai Hit). To realize quick thermal equilibrium of the sample, the coverslip was in tight contact with a copper plate and the thermoplate, as shown in Fig. 2(a). The sample temperature was monitored by a data logger (TR-71wb, T&D) via a thermistor (TR-0106, T&D) embedded in the copper plate. After the preset temperatures of the thermoplate and lens heater were changed, the monitored temperature stabilized within 30 min with a fluctuation of 0.1 °C. Thus, we assumed that the sample temperature was the monitored temperature after thermal equilibrium. Table S1 shows the relation between the preset and sample temperatures. Although the sample temperature did not reach the preset temperature due to heat dissipation, the sample temperature exhibited a linear increase with preset temperature.

**Table S1.** Relation between preset temperatures of thermoplate and lens heater and sample temperature.

| Preset temperatures [°C] | 22.0 | 26.0 | 30.0 | 34.0 | 38.0 | 42.0 |
|---|---|---|---|---|---|---|
| Sample temperature [°C] | 22.0 | 25.7 | 29.4 | 33.1 | 36.8 | 40.5 |

**PL spectrum analysis.**

Because the PL spectrum of SiV-DNDs includes not only the SiV center's sharp ZPL but also the DND-originated broad spectrum, we fitted the spectrum to a sum of Lorentzian and linear background terms:

$$f(\lambda) = A \frac{\gamma/\pi}{\left(\lambda - \lambda_\mathrm{p}\right)^2 + \gamma^2} + B\left(\lambda - \lambda_\mathrm{p}\right) + C,$$

where $\lambda_\mathrm{p}$ is the ZPL peak wavelength, $2\gamma$ is the spectral width (FWHM), and $A$ - $C$ are the weighting factors. The fitting wavelength range was fixed from 732 to 742 nm for all data. We obtained the slope $\Delta\lambda/\Delta T$ from the experimental data ($\lambda_\mathrm{p}$ vs. sample temperature) fitted to a linear function with $R^2 > 0.85$.

**Histograms of the peak wavelength, the spectral width, and the temperature sensitivity.**

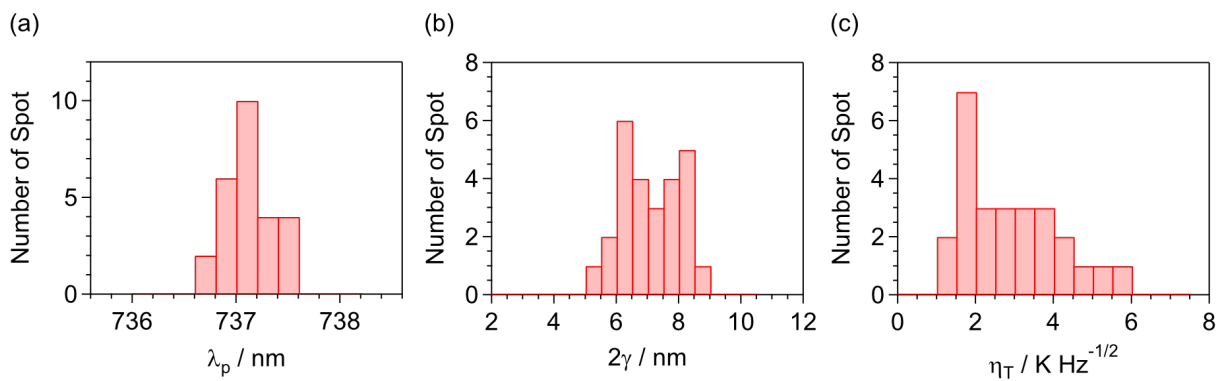

**Figure S1.** Histograms of (a) ZPL peak wavelength, (b) ZPL peak width (FWHM), and (c) temperature sensitivity for 26 bright spots. The average values are 737.1 ± 0. 2 nm, 7.1 ± 1.0 nm, and 2.9 ± 1.3 K Hz$^{-1/2}$, respectively. All data were measured at a sample temperature of 22.0 °C.

Figure 4(d) showed a histogram of the ZPL peak sensitivity to temperature ($\Delta\lambda/\Delta T$) for 26 bright spots. Here, we show the histograms for other parameters. Figures S1(a) and S1(b) show the histograms for the ZPL peak wavelength and the spectral width, respectively. The peak wavelength is distributed in a narrow region around 737 nm, and the spectral width is distributed around 7 nm. These results indicate that the SiV center PL spectrum of a bright spot is comparable to that for bulk diamond or small-strain nanodiamonds.[1,2] Figure S1(c) shows a histogram of the temperature sensitivity ($\eta_T$). It should be noted that the temperature sensitivity ($\eta_T$) depends not only on the peak sensitivity to temperature ($\Delta\lambda/\Delta T$) but also on the number of SiV centers. The sensitivity peaks at around 1.5 K/$\sqrt{\text{Hz}}$, but is widely distributed compared to the peak sensitivity to temperature ($\Delta\lambda/\Delta T$) in Fig. 4(c), which is considered to be due to the difference in the number of SiV centers at the measured spots and/or differences in the surrounding environment of each bright spot (aggregated DNDs). Note that there is no clear correlation between the peak wavelength or the spectral width and the slope $\Delta\lambda/\Delta T$.

**Estimation of the number of SiV centers in a bright spot.**

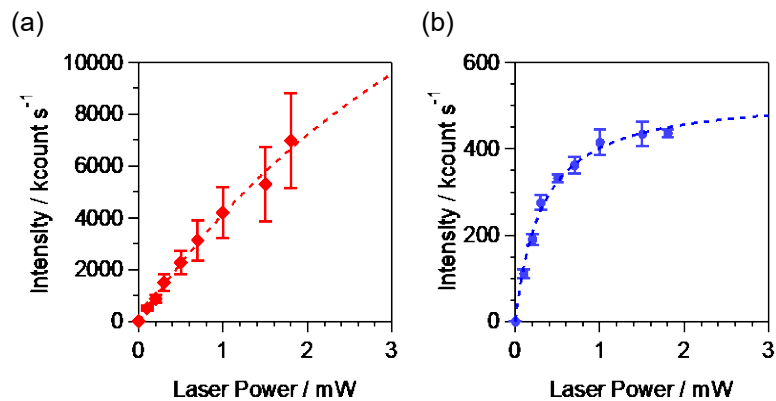

**Figure S2.** Saturation curves for (a) SiV centers extracted from bright spots and (b) single NV centers in CVD diamond substrate. The dotted line represents the best fit to a saturation curve function.

We measured the saturation curves and estimated the number of SiV centers for the bright spots in a confocal image of SiV-DNDs. First, we extracted the SiV center signal as the difference in PL intensity between a bright spot and a region without bright spots, as shown in Figs. 2(c) and 2(d). Figure S2(a) shows the saturation curve averaged for six bright spots. From these data, the PL intensity is 4200 ± 1000 kcps at a laser power of 1 mW. The dotted line represents the fitting result for the saturation curve function $I(w) = wI_{sat} / (w + w_{sat})$, where $I$ is the PL intensity, $w$ is the laser power, $I_{sat}$ is the PL intensity in saturation, and $w_{sat}$ is the laser power in saturation. Although the maximum laser power is limited to 1.8 mW for our optical system, we can estimate $I_{sat}$ to 28000 ± 9000 kcps and $w_{sat}$ to 5.7 ± 2.2 mW. We also measured the saturation curves for single NV centers in a CVD diamond substrate for comparison. Figure S2(b) shows the results averaged for four single NV centers. The PL intensity is 420 ± 30 kcps at a laser power of 1 mW, and $I_{sat}$ and $w_{sat}$ are estimated to be 530 ± 20 kcps and 0.32 ± 0.03 mW, respectively.

Although $I_{sat}$ and $w_{sat}$ depend on not only the material properties but also the detection efficiency of the optical system (numerical aperture of the objective, pinhole diameter, detection filter type, etc.), the features of the curve are comparable to previous results.[2-4] In addition, the saturation curves reported by Rogers et al.[4] indicate that the PL intensity for the single SiV center at 1 mW is about 1/6 of that of a single NV center for a given optical system. Hence, we can assume that the PL intensity for a single SiV center is 70 kcps at 1 mW. From a comparison of the PL intensity for an SiV center at each bright spot with that for a single SiV center, we can estimate that each bright spot has 60 ± 15 SiV centers in a focal spot.

**Diamond quantum thermometry.**

A diamond quantum sensor is a potential high-sensitive nanoscale thermometer that utilizes the optical and electrical spin properties of color defect centers in diamond. Here, we summarized the recently reported temperature sensitivities of the color defect centers in diamond with respect to the diamond particle size, as shown in Figure S3 and Table S2. The sensitivity can not be straightforwardly compared because it depends not only on the size but also on the types of the thermometry method.[5] Therefore, they are also shown in Figure S3 and Table S2 for comparison. Concerning the size of the nanodiamonds, in our research, the polyglycerol (PG)-coating was applied to disperse the DND in water, which is an essential treatment for biological application and increased the particle size of DND from 10.8 nm to 20 nm as described in the main text. Note, in Figure S3, we plot both of them since the PG coating will not deteriorate the performance of the temperature sensitivity.

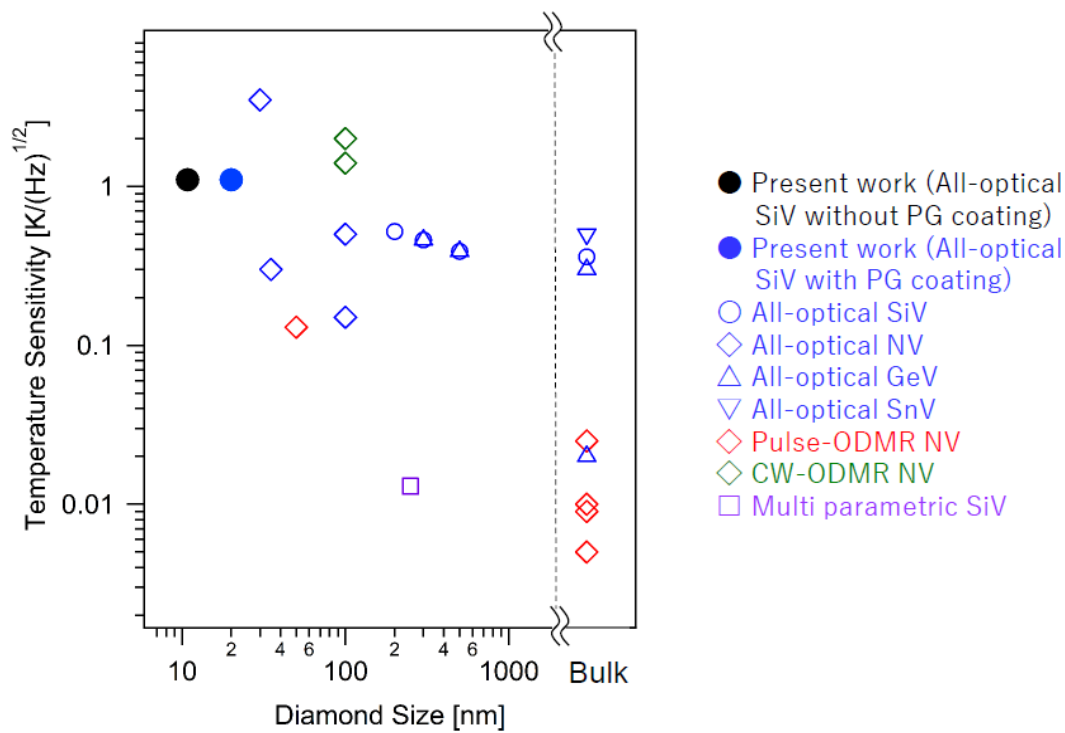

**Figure S3.** Plots of temperature sensitivity versus diamond size; data are summarized in table S2 with references. The black filled circle (●) and the filled blue circle (●) show the present works without and with PG coating, respectively. The open blue circle (○), the blue rhombus (◇), the blue trigonometric (△), the blue inverse trigonometric (▽) show all-optical thermometries of the SiV center, the NV center, the GeV center, and the SnV center, respectively. The red rhombus (◇) and the green rhombus (◇) show the NV center thermometry by pulse-ODMR method and by CW-ODMR method, respectively. The purple scare (□) shows the result of multiparametric data analysis in the SiV center.

**Table S2.** Size and sensitivity of diamond thermometers.

| Defect | Form | Sensor size (nm) | Method | Sensitivity (mK/$\sqrt{\text{Hz}}$) | Ref. |
|---|---|---|---|---|---|
| Spin-based thermometry | | | | | |
| NV | Bulk | - | Pulse-ODMR | 25 | [6] |
| NV | Bulk | - | Pulse-ODMR | 9 | [7] |
| NV | Bulk | - | Pulse-ODMR | 5 | [8] |
| NV | Bulk | - | Pulse-ODMR | 10 | [9] |
| NV | ND | 50 | Pulse-ODMR | $1.3\times10^2$ | [8] |
| NV | ND | 100 | CW ODMR | $2.0\times10^3$ | [10] |
| NV | ND | 100 | CW ODMR | $1.4\times10^3$ | [11] |
| All-optical thermometry | | | | | |
| SiV (Our result) | ND | 20 | ZPL shift | $1.1\times10^3$ | - |
| SiV | Bulk | - | ZPL shift | $3.6\times10^2$ | [12] |
| GeV | Bulk | - | ZPL intensity, shift | $3.0\times10^2$ | [13] |
| GeV | Bulk | - | Fiber optic ZPL intensity, shift | 20 | [14] |
| SnV | Bulk | - | ZPL linewidth, shift | $5.0\times10^2$ | [15] |
| NV | ND | 35 | ZPL intensity | $3.0\times10^2$ | [16] |
| NV | ND | 30 | Debye-Waller factor | $3.5\times10^3$ | [17] |
| NV | ND | 100 | ZPL shift | $2.0\times10^3$ | [18] |
| NV | ND | 100 | ZPL intensity | $1.5\times10^2$ | [19] |
| NV | ND | 100 | ZPL shift | $4.6\times10^2$ | [19] |
| SiV | ND | 200 | ZPL intensity | $5.2\times10^2$ | [12] |
| GeV | ND | 300 | Antistokes | $4.6\times10^2$ | [20] |
| SiV / GeV | ND | 500 | Fluorescence ratio | $3.9\times10^2$ | [21] |
| SiV | ND | 250 | multiparametric | 13 | [22] |

# References

1       Nguyen, C. T.; Evans, R. E.; Sipahigil, A.; Bhaskar, M. K.; Sukachev, D. D.; Agafonov, V. N.; Davydov, V. A.; Kulikova, L. F.; Jelezko, F.; Lukin, M. D. All-Optical Nanoscale Thermometry with Silicon-Vacancy Centers in Diamond. *Appl. Phys. Lett.* **2018**, *112*, 203102.

2       Bolshedvorskii, S. V.; Zeleneev, A. I.; Vorobyov, V. V.; Soshenko, V. V.; Rubinas, O. R.; Zhulikov, L. A.; Pivovarov, P. A.; Sorokin, V. N.; Smolyaninov, A. N.; Kulikova, L. F.; Garanina, A. S.; Lyapin, S. G.; Agafonov, V. N.; Uzbekov, R. E.; Davydov, V. A.; Akimov, A. V. Single Silicon Vacancy Centers in 10 nm Diamonds for Quantum Information Applications. *ACS Appl. Nano Mater.* **2019**, *2*, 4765–4772.

3       Bolshedvorskii, S. V.; Vorobyov, V. V.; Soshenko, V. V.; Shershulin, V. A.; Javadzade, J.; Zeleneev, A. I.; Komrakova, S. A.; Sorokin, V. N.; Belobrov, P. I.; Smolyaninov, A. N.; Akimov, A. V. Single Bright NV Centers in Aggregates of Detonation Nanodiamonds. *Opt. Mater. Express* **2017**, *7*, 4038.

4       Rogers, L. J.; Jahnke, K. D.; Teraji, T.; Marseglia, L.; Müller, C.; Naydenov, B.; Schauffert, H.; Kranz, C.; Isoya, J.; McGuinness, L. P.; Jelezko, F. Multiple Intrinsically Identical Single-Photon Emitters in the Solid State. *Nat. Commun.* **2014**, *5*, 4739.

5       Fujiwara, M.; Shikano, Y. Diamond quantum thermometry: from foundations to applications, *Nanotechnology*, **2021** 32 482002.

6       Toyli, D. M.; Casas, C. F.; Christle, D. J.; Dobrovitski, V. V.; Awschalom, D. D. Fluorescence thermometry enhanced by the quantum coherence of single spins in diamond. *Proc. Natl. Acad. Sci.* **2013**, 110, 8417-8421.

7       Kucsko, G.; Maurer, P. C.; Yao, N. Y.; Kubo, M.; Noh, H. J.; Lo, P. K.; Park, H.; Lukin, M. D. Nanometre-Scale Thermometry in a Living Cell. *Nature* **2013**, *500*, 54–58.

8       Neumann, P.; Jakobi, I.; Dolde, F.; Burk, C.; Reuter, R.; Waldherr, G.; Honert, J.; Wolf, T.; Brunner, A.; Shim, J. H.; Suter, D.; Sumiya, H.; Isoya, J.; Wrachtrup, J. High-Precision Nanoscale Temperature Sensing Using Single Defects in Diamond. *Nano Lett.* **2013**, 13, 2738-2742.

9       Wang, J.; Feng, F.; Zhang, J.; Chen, J.; Zheng, Z.; Guo, L.; Zhang, W.; Song, X.; Guo, G.; Fan, L.; Zou, C.; Lou, L.; Zhu, W.; Wang, G. High-sensitivity temperature sensing using an implanted single nitrogen-vacancy center array in diamond. *Phys. Rev. B* **2015**, 91, 155404.

10      Tzeng, Y. K.; Tsai, P. C.; Liu, H. Y.; Chen, O. Y.; Hsu, H.; Yee, F. G.; Chang, M. S.; Chang, H. C. Time-Resolved Luminescence Nanothermometry with Nitrogen-Vacancy Centers in Nanodiamonds. *Nano Lett.* **2015**, 15, 3945-3952.

11      Fujiwara, M.; Sun, S.; Dohms, A.; Nishimura, Y.; Suto, K.; Takezawa, Y.; Oshimi, K.; Zhao, L.; Sadzak, N.; Umehara, Y.; Teki, Y.; Komatsu, N.; Benson, O.; Shikano, Y.; Kage-Nakadai, E. Real-Time Nanodiamond Thermometry Probing in-Vivo Thermogenic Responses. *Sci. Adv.* **2020**, *6,* No. eaba9636.

12      Nguyen, C. T.; Evans, R. E.; Sipahigil, A.; Bhaskar, M. K.; Sukachev, D. D.; Agafonov, V. N.; Davydov, V. A.; Kulikova, L. F.; Jelezko, F.; Lukin, M. D. All-Optical Nanoscale Thermometry with Silicon-Vacancy Centers in Diamond. *Appl. Phys. Lett.* **2018**, *112*, 203102.

13      Fan, J. W.; Cojocaru, I.; Becker, J.; Fedotov, I. V.; Alkahtani, M. H. A.; Alajlan, A.; Blakley, S.; Rezaee, M.; Lyamkina, A.; Palyanov, Y. N.; Borzdov, Y. M.; Yang, Y. P.;

Zheltikov, A.; Hemmer, P.; Akimov, A. V. Germanium-Vacancy Color Center in Diamond as a Temperature Sensor. *ACS Photonics* **2018**, 5, 765-770.

14    Blakley, S.; Liu, X.; Fedotov, I.; Cojocaru, I.; Vincent, C.; Alkahtani, M.; Becker, J.; Kieschnick, M.; Lühman, T.; Meijer, J.; Hemmer, P.; Akimov, A.; Scully, M.; Zheltikov, A. Fiber-Optic Quantum Thermometry with Germanium-Vacancy Centers in Diamond. *ACS Photonics* **2019**, 6,1690-1693.

15    Alkahtani, M.; Cojocaru, I.; Liu, X.; Herzig, T.; Meijer, J.; Küpper, J.; Lühmann, T.; Akimov, A. V.; Hemmer, P. R. Tin-Vacancy in Diamonds for Luminescent Thermometry. *Appl. Phys. Lett.* **2018**, *112*, 241902.

16    Plakhotnik, T.; Aman, H.; Chang, H. C. All-optical single-nanoparticle ratiometric thermometry with a noise or of 0.3 K Hz^-1/2. *Nanotechnology* **2015**, 26, 245501.

17    Plakhotnik, T.; Doherty, M. W.; Cole, J. H.; Chapman, R.; Manson, N. B. All-Optical Thermometry and Thermal Properties of the Optically Detected Spin Resonances of the NV- Center in Nanodiamond. *Nano Lett.* **2014**, *14*, 4989–4996.

18    Tsai, P. C.; Epperla, C. P.; Huang, J. S.; Chen, O. Y.; Wu, C. C.; Chang, H. C. Measuring Nanoscale Thermostability of Cell Membranes with Single Gold–Diamond Nanohybrids. *Angew. Chem. Int. Ed.* **2017**, 56, 3025-3030.

19    Hui, Y. Y.; Chen, O. Y.; Azuma, T.; Chang, B. M.; Hsieh, F. J.; Chang, H. C. All-Optical Thermometry with Nitrogen-Vacancy Centers in Nanodiamond-Embedded Polymer Films. *J. Phys. Chem. C* **2019**, 123, 15366-15374.

20    Tran, T. T.; Regan, B.; Ekimov, E. A.; Mu, Z.; Zhou, Y.; Gao, W.; Narang, P.; Solntsev, A. S.; Toth, M.; Aharonovich, I.; Bradac, C. Anti-Stokes excitation of solid-state quantum emitters for nanoscale thermometry. *Sci. Adv.* **2019**, eaav9180.

21    Chen, Y.; Li, C.; Yang, T.; Ekimov, E. A.; Bradac, C.; Toth, M.; Aharonovich, I.; Tran, T. T. Real-time ratiometric optical nanoscale thermometry. *arXiv*:2112.01758, **2021**.

22    Choi, S.; Agafonov, V. N.; Davydov, V. A.; Plakhotnik, T. Ultrasensitive All-Optical Thermometry Using Nanodiamonds wi th a High Concentration of Silicon-Vacancy Centers and Multiparametric Data Analysis. *ACS Photonics* **2019**, *6*, 1387–1392.